\documentclass[10pt]{elsarticle}
\usepackage{epsfig}

\begin{document}

\title{Schwinger - Dyson equation and NJL approximation in massive gauge theory with fermions}

\author[UWO,ITEP]{M.A.~Zubkov\footnote{Corresponding author,
 e-mail: zubkov@itep.ru}}

\address[UWO]{University of Western Ontario,  London, ON, Canada N6A 5B7 }

\address[ITEP]{ITEP, B.Cheremushkinskaya 25, Moscow, 117259, Russia
}

\begin{abstract}
We consider massive $SU(N)$ gauge theory  with fermions. Gauge bosons become massive due to the interaction with the scalar field, whose vacuum average provides the spontaneous breakdown of gauge symmetry. We investigate Dyson - Schwinger equation for the fermion propagator written in ladder approximation and in Landau gauge. Our analysis demonstrates that the chiral symmetry breaking in the considered theory is the strong coupling phenomenon. There are the indications that there appears the second order phase transition  between chirally broken and symmetric phases of the theory at the value of coupling constant $\alpha_c = (1+\gamma)\times \frac{\pi}{3}\times \frac{1}{2 C_2(F)}$, where $0<\gamma<1$, and $\gamma$ depends on the scale, at which the fluctuations of the scalar field destroy the gauge boson mass. In the broken phase near the critical value of $\alpha$ the Dyson - Schwinger equation is approximated well by the gap equation of the effective Nambu - Joina - Lasinio model with the value of cutoff around gauge boson mass $M$ and the effective four - fermion coupling constant $\frac{4 \pi \alpha}{M^2}\times \frac{2C_2(F)}{N}$. The dynamical fermion mass $m$ may be essentially smaller than $M$.
\end{abstract}



\maketitle

\section{Introduction}

The Nambu - Jona - Lasinio (NJL) approximation in field theory is the approximation with the effective 4 - fermion interaction \cite{NJL}. It allows to understand qualitatively formation of fermion condensates in various physical systems \cite{Miransky:1994vk} from superconductivity and superfluidity \cite{Nambu} to top quark condensation \cite{Miransky}. Any NJL model is the non - renormalizable low energy approximation to the  microscopic theory. Therefore, its predictions depend strongly on the way it is regularized. In many publications on NJL models the ordinary cutoff regularization was assumed, in which the cutoff $\Lambda$ is present in all loop integrals. In most of such papers physical quantities are evaluated in one - loop approximation (i.e. in the leading order in $1/N$ expansion). The higher loops were simply disregarded. The examples are the NJL approximation to QCD \cite{NJLQCD}, Technicolor \cite{Simmons}, the papers on the models of top quark condensation (TC) \cite{Miransky,topcolor1}, on the Extended Technicolor (ETC) \cite{Simmons}, on the topcolor assisted technicolor \cite{Hill:1994hp,LaneEichten,Popovic:1998vb, Braam:2007pm}, on the top - seesaw \cite{topseesaw}. The use of the one - loop approximation may cause a confusion because formally the contributions of higher loops to various physical quantities are strong. In \cite{cveticreview,cvetic} it has been shown that the next to leading (NTL) order approximation to the  fermion mass $M_f$ is weak compared to the one - loop approximation only if $M_f \sim \Lambda$. It follows from analytical results and from numerical simulations made within the lattice regularization \cite{latticeNJL} (in which lattice spacing is related to the cutoff as $a \sim \frac{\pi}{\Lambda}$) that the dimensional physical quantities in the relativistic NJL models are typically of the order of the cutoff unless their small values  are protected by symmetry. As a result, strictly speaking, the one - loop approximation to the NJL model in cutoff regularization may give a more or less reasonable contribution to the physical quantities only if those quantities are of the order of the cutoff. This is the case of QCD and conventional technicolor.

At the same time, for example, in the case of the models of top - quark condensation, top seesaw, and ETC, formally the one - loop results cannot be used because the cutoff is assumed to be many orders of magnitude larger than the generated fermion mass. That means, that in order to use the one - loop results in the mentioned cases in the NJL model with cutoff regularization we should start from the action of the model with the additional counter - terms. If these counter - terms cancel quadratic divergences in the next to leading orders of $1/N$ expansion, then this expansion may be applied and, in particular, the one - loop results give reasonable estimates to the physical quantities. Such a redefined NJL model is actually equivalent to the original NJL model defined in zeta or dimensional regularization. The four fermion coupling constants of the two regularizations are related by the finite renormalization (see \cite{ZetaNeutrino}, Appendix, Sect. 4.2.).
The NJL models in zeta regularization were considered, for example, in \cite{ZetaNeutrino,ZetaHiggs}. The NJL model in dimensional regularization was considered in \cite{NJLdim}.

It is widely believed that there is the exchange by massive gauge bosons behind the NJL models of top quark condensation, top seesaw, and ETC. If so, the appearance of the one - loop gap equation of NJL model should follow from the direct investigation of the theory with massive gauge fields interacting with fermions. The obvious difficulty for establishing this result is related to the fact, that chiral symmetry breaking in the considered theory may appear to be a strong coupling effect. (Indeed this will be explicitly demonstrated in the present paper.) The applicability of various analytical methods to the investigation of strongly coupled theories is limited. Actually, all results that may be obtained here are to be verified by direct lattice numerical simulations. Nevertheless, there exist the analytical methods, which were applied successfully to certain strongly coupled theories, although without the rigorous proof that this may be done. One of such methods is the use of truncated Schwinger - Dyson equation (for the review see \cite{Fomin:1984tv,Roberts:1994dr,Alkofer:2000wg,Fischer:2006ub,Bashir:2012fs,Roberts:2007ji} and references therein).

In the present paper we apply the Schwinger - Dyson (SD) equation with the kernel that corresponds to the single exchange by  massive gauge boson and bare interaction vertex. For the consideration of this problem for the massless $U(1)$ gauge bosons see \cite{MiranskyGap} and also \cite{Roberts:1994dr,Miransky:1994vk}. The output of that study is the prediction that there may exist the phase transition in QED at a rather strong value of coupling constant $\alpha_{c1} \approx \pi/3$. For $\alpha>\alpha_{c1}$ the theory may appear to be in the phase with broken chiral symmetry. The value of dynamical fermion mass $m$ should be related to the only massive parameter existing in the theory that is the ultraviolet cutoff $E_{UV}$. The relation between them is given by the so - called Miransky scaling $m \approx 4 E_{UV} \, {\rm exp}\Big(-\frac{\pi }{\sqrt{\frac{\alpha}{\alpha_{c1}}-1}}\Big)$. The existence of the phase transition has been confirmed later by lattice numerical investigations \cite{latticeQEDrev,latticeQED,latticeQED2,Gockeler:1989wj}. However, in compact lattice QED this is the first order phase transition \cite{latticeQED}. In non - compact formulation \cite{latticeQED2} the lattice study indicates the second order phase transition but the above scaling has not been confirmed \cite{Gockeler:1989wj}. Unlike the more complicated models the analysis of QED is exhaustive. The output of this study is that Schwinger - Dyson equation in Landau gauge indeed gives reasonable qualitative pattern of chiral symmetry breaking. However, its quantitative predictions should be taken with care.

The similar problem has been studied in massive electrodynamics using Schwinger - Dyson equation (see \cite{Kondo:1989vs} and references therein). Later this approach was used as a building block for the construction of the theories with dynamical symmetry breaking \cite{Yoshida:1995ve}. However, the gauge different from the Landau gauge was used, and the kernels of the equations are so complicated that their direct analysis was not performed. Instead, the kernels were substituted by some approximating expressions without strong justification. Besides, the wave function renormalization function $A(z)$ was taken equal to unity as for the massless case although the corresponding kernel does not vanish.  The output and method of \cite{Kondo:1989vs} is, in general, similar to that of the study of pure QED of \cite{MiranskyGap}. However, the approximations used require the re - check using another technique. This will be done in the present paper.

We shall explore Dyson - Schwinger equation in Landau gauge, and use the kernels as they are. The result for the critical coupling constant is, in principle, similar to that of \cite{Kondo:1989vs}. In the massive gauge theory there exists the finite dimensional parameter - the gauge boson mass $M$. It should be related with the dynamical fermion mass $m$. The Dyson - Schwinger equation in Landau gauge does not contain ultraviolet divergences. We assume, that the gauge boson becomes massive  because of the interaction with the (possibly, composite) scalar field. This scalar field is condensed, and gauge boson mass appears as a result of Higgs mechanism. In this pattern the finite ultraviolet cutoff $E_{UV} \gg M$ appears that marks the scale at which the fluctuations of the mentioned scalar field wash out the gauge boson mass.

We shall derive gap equation near to the criticality (which has not been not done in \cite{Kondo:1989vs}) and observe, that its form coincides {\it exactly} with the one loop gap equation of the Nambu - Jona - Lasinio model with the suitable definition of the four - fermion coupling constant and the cutoff of the order of the gauge boson mass. We present our results in the form that is valid for a wide variety of gauge groups and fermion representations.

It is worth mentioning that the numerical lattice study of the system with massive gauge bosons coupled to fermions has been performed both in the case of $U(1)$ gauge group and $SU(2)$ gauge group. In both cases there are the indications of the continuous phase transition. In case of $U(1)$ gauge theory \cite{Franzki:1995xb} the evidence is given that the model may indeed be approximated at strong enough values of gauge coupling constant by the the effective one - loop NJL model understood through the one - loop results. However, in this case (see \cite{U1lattice} and references therein) the formal continuum limit is, most likely non - interacting because of the Landau pole. In the $SU(2)$ case (see, for example, \cite{SU2lattice} and references therein) the continuum limit of the lattice model has not been investigated and, therefore, the value of continuum critical coupling constant has not been calculated. Also the question of the possibility to approximate the model by the one - loop NJL results has not been investigated.



\section{Dyson - Schwinger equation in rainbow quenched approximation}

Let us consider the system of fermion $\psi$ and scalar $\Phi$ coupled to gauge field $A_{\mu}$ (with the field strength $A_{\mu\nu}$).  $A$ belongs to the group $G = SU(N)$. We assume that $\psi$ belongs to the fundamental representation of this group while $\Phi$ is the $N\times N$ matrix field, and group $G$ acts on its first index. The action of the system has the form:
\begin{equation}
S = \int d^4 x \Bigl(-\frac{1}{2 g^2} {\rm Tr} A^2_{\mu\nu} + i\bar{\psi} \gamma D \psi + {\rm Tr}\,[D\Phi]^+ D\Phi  - V\Big( \Phi^+ \Phi \Big)\Bigr)\label{S0}
\end{equation}
We assume, that $G$ is broken spontaneously due to the potential $V$, which gives rise to vacuum average  $\langle\Phi\rangle = \frac{M}{g}\times {\bf 1}$. "Angular" modes of the field $\Phi$ are eaten by the gauge bosons while the "radial" modes are massive with mass matrix $\hat{m}_{\Phi}$ that depends on the details of $V$. At the energies much smaller than the eigenvalues of $\hat{m}_{\Phi}$ we arrive at the effective action:
\begin{equation}
S = \int d^4 x \Bigl(-\frac{1}{2 g^2} {\rm Tr} A^2_{\mu\nu} + i\bar{\psi} \gamma D \psi + \frac{M^2}{g^2}{\rm Tr} A^2\Bigr)\label{S01}
\end{equation}
Notice, that this effective action may be valid up to the energies much larger than $M$ if all eigenvalues of $\hat{m}_{\Phi}$ are sufficiently large.
Further we shall consider the system with effective action Eq. (\ref{S01}) in Landau gauge.
The inverse fermion propagator $D^{-1}$ in Euclidean space - time in ladder approximation and with the contributions of ghosts (that appear in higher loops) neglected satisfies Dyson - Schwinger equation
\begin{equation}
D^{-1}(p) = A(p^2) \gamma p - i B(p^2) = \gamma p - i \Sigma(p),
\end{equation}
where $\Sigma(p)$ is the self energy operator
\begin{eqnarray}
i\Sigma(p) &=& g^2 \int \frac{d^4 k}{(2\pi)^4} \gamma_{\mu} T^a\frac{1}{A(k^2)\gamma k - i B(k^2)} \nonumber\\&&\frac{g^{\mu\nu} - \frac{{(p-k)}^{\mu}{(p-k)}^{\nu}}{(p-k)^2}}{{(p-k)}^2 + M^2}T^a\gamma_{\nu}\nonumber\\
&=& 3 g^2 C_2(F) i \int \frac{d^4 k}{(2\pi)^4} \frac{B(k^2)}{A^2(k^2) k^2 +  B^2(k^2)} \nonumber\\&&\frac{1}{{(p-k)}^2 + M^2}\nonumber\\&&- g^2 C_2(F) \gamma p \frac{1}{p^2}  \int \frac{d^4 k}{(2\pi)^4}\frac{A(k)}{A^2(k^2) k^2 +  B^2(k^2)} \nonumber\\&&\frac{kp + 2 \frac{(p-k,p)(p-k,k)}{(p-k)^2}}{{(p-k)}^2 + M^2}\nonumber
\end{eqnarray}
Here $T^aT^a = C_2(F) \times 1$. We obtain the system of equations for the functions $A$ and $B$:
\begin{eqnarray}
B(p^2) &=& 3 \frac{\alpha}{4\pi} 2C_2(F)  \int \frac{d^4 k}{\pi^2} \frac{B(k^2)}{A^2(k^2) k^2 +  B^2(k^2)}\nonumber\\&& \frac{1}{{(p-k)}^2 + M^2}\nonumber\\
A(p^2) & = & 1 + \frac{\alpha}{4\pi} 2C_2(F) \frac{1}{p^2}  \int \frac{d^4 k}{\pi^2} \frac{A(k)}{A^2(k^2) k^2 +  B^2(k^2)} \nonumber\\&&\frac{kp + 2 \frac{(p-k,p)(p-k,k)}{(p-k)^2}}{{(p-k)}^2 + M^2}\nonumber
\end{eqnarray}

Let us introduce the notation $m(k^2) = B(k^2)/A(k^2)$. We may rewrite equations in terms of functions $A$ and $m$:
\begin{eqnarray}
A(p^2)m(p^2) &=& 3 \frac{\alpha}{4\pi} 2C_2(F)  \int \frac{d^4 k}{\pi^2} \frac{A^{-1}(k^2) m(k^2)}{k^2 +  m^2(k^2)} \nonumber\\&& \frac{1}{{(p-k)}^2 + M^2}\nonumber\\
A(p^2) & = & 1 + \frac{\alpha}{4\pi} 2C_2(F) \frac{1}{p^2}  \int \frac{d^4 k}{\pi^2} \frac{A^{-1}(k)}{ k^2 +  m^2(k^2)} \nonumber\\&& \frac{kp + 2 \frac{(p-k,p)(p-k,k)}{(p-k)^2}}{{(p-k)}^2 + M^2}\label{redgapA}
\end{eqnarray}
The integration over angles leads to
\begin{eqnarray}
A(u)m(u) &=& 3 \frac{\alpha}{4\pi} 2C_2(F)  \int {d z} \frac{A^{-1}(z)m(z)}{z +  m^2(z)} K(z,u)\label{gapeq}\\
A(u) &=& 1 + \frac{\alpha}{4\pi} 2C_2(F)  \int {d z} A^{-1}(z)\frac{1}{z +  m^2(z)} L(z,u),\nonumber
\end{eqnarray}
where  $u = p^2 $. The kernels $K$ and $L$ are given by
\begin{eqnarray}
K(z,u) &=& \frac{1}{2 u} \Big(u+z+M^2-\sqrt{(u+z+M^2)^2-4 uz} \Big)\nonumber\\
L(z,u) &=& \frac{1}{4M^2 u^2} \Big[3\Big(u+z+M^2\Big)M^2\Big(u+z+M^2\nonumber\\&&-\sqrt{(u+z+M^2)^2-4 uz} \Big)\nonumber\\&&
-\Big((u+z+M^2)^3-\Big((u+z+M^2)^2-4 uz\Big)^{3/2} \Big)\nonumber\\&&
+\Big((u+z)^3-\Big((u+z)^2-4 uz\Big)^{3/2} \Big)\Big]\label{kernels}
\end{eqnarray}



\section{Evaluation of $A(z)$}

\begin{figure}
\begin{center}
 \epsfig{figure=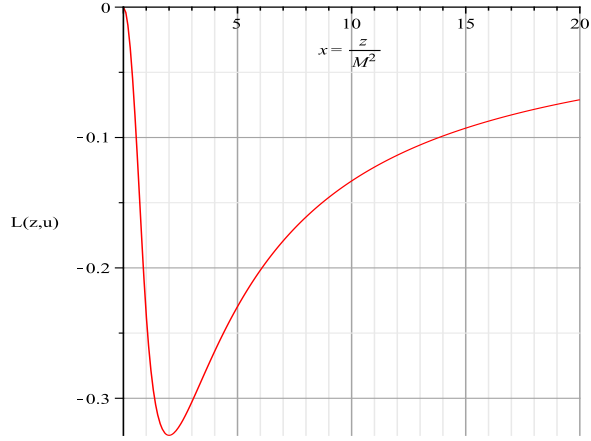,height=60mm,width=80mm,angle=0}
\caption{\label{fig12} Kernel $L(z,u)$ as a function of $x = z/M^2$ for $u = M^2$.    }
\end{center}
\end{figure}

In order to evaluate function $A(z)$ near the critical value of $\alpha$, where the nontrivial solution $m\ne 0$ appears we set  $m = 0$ in the second equation of Eq. (\ref{gapeq}). This gives
\begin{eqnarray}
A(u) &=& 1 + \frac{1}{\xi}  \int {d z} A^{-1}(z)\frac{1}{z} L(z,u)\label{gapeqL}
\end{eqnarray}
Here we denote
\begin{equation}
\xi = \frac{4\pi}{2\alpha  C_2(F)}
\end{equation}
As an example, we represent the Kernel $L(z,u)$ for $u = M^2$ as a function of $z$ in Fig. \ref{fig12}.
Notice, that the kernel $L(z,u)$ has an asymptotic form $L(z,u) \approx 0$ at $u \gg M^2$. Therefore, function $A(z)$ should tend to $1$ at $z \rightarrow \infty$. Next, we shall see in the next sections, that near to the criticality the value of $1/\xi$ is of the order of $0.1$. Therefore, we are able to solve Eq. (\ref{gapeqL}) perturbatively. Our zero approximation is $A^{(0)}(z) = 1$. The first approximation is given by
\begin{equation}
A(u) = 1 + \frac{g(u)}{\xi},\label{Afin0}
\end{equation}
where
\begin{eqnarray}
g(u) &=&  \int \frac{dz}{z} L(z,u)\label{gapeqLg}
\end{eqnarray}
The integral may be calculated explicitly (we use MAPLE package):
\begin{eqnarray}
g(u) &=& (1/24) (-6\, {\rm arcsh}((1/2)\,(x-1)/\sqrt{x})\,x^3\nonumber\\&&+12\,x^3\,{\rm log}(x+1)-12\,x^3\,{\rm log}(x)-12\,x^2\nonumber\\&&-36\,x\,{\rm log}(x+1)+18\,{\rm log}(x)\,x+24\,x\nonumber\\&&-18\,x\,{\rm arcsh}((1/2)\,(x-1)/\sqrt{x})\nonumber\\&&-12\,{\rm arcsh}((1/2)\,(x-1)/\sqrt{x})-24\,{\rm log}(x+1)\nonumber\\&&+12\,{\rm log}(x)+6\,{\rm arcth}((x-1)/(x+1))\,x^3\nonumber\\&&-18\,{\rm arcth}((x-1)/(x+1))\,x\nonumber\\&&-12\,{\rm arcth}((x-1)/(x+1)))/x^2, \label{g}
\end{eqnarray}
where $x = u/M^2$.
\begin{figure}
\begin{center}
 \epsfig{figure=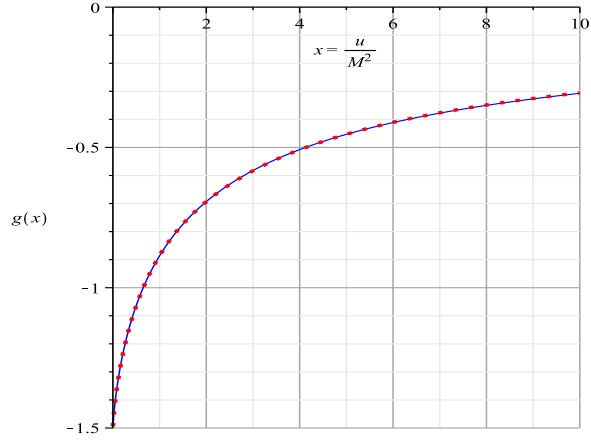,height=60mm,width=80mm,angle=0}
\caption{\label{fig3} Function $g(u)$ (solid line) and our fit of Eq. (\ref{fitA}) (dotted line).    }
\end{center}
\end{figure}
The asymptotic expressions for the function $g(x)$ are:
\begin{eqnarray}
g(u) &\approx & -3/2+(-(1/2)\, {\rm log}(x)+5/12)\, x, \quad x = \frac{u}{M^2} \ll 1\nonumber\\
g(u)& \approx &-\frac{3}{2x}{\rm log}(x) - \frac{1}{2x}\,\quad x \gg 1
\end{eqnarray}
We are able to interpolate between the two:
\begin{equation}
g(u) \approx \frac{-\frac{3}{2} + (\frac{5}{12}-\frac{1}{2}\, {\rm log}(x))\, x}{1 + \frac{1}{3}\, x^2}-\frac{1}{2}\,\frac{ x^2}{1.62491403813569+4.23237714074304\, x+x^3} \label{fitA}
\end{equation}
In Fig. \ref{fig3} we represent both Eq. (\ref{g}) and the fit of Eq. (\ref{fitA}). One can see, that our fit approximates the form of $g(u)$ reasonably well at all values of $u$. In practise we do not observe the difference between the two on the plot.

In order to check the accuracy of our solution we estimate the next iteration for the value of $A(0)$. We have $A^{(2)}(0) = 1 + \frac{1}{\xi} \int \frac{dz}{z A^{(1)}(z)}L(z,0)$ while $A^{(1)}(0) = 1 + \frac{1}{\xi} \int \frac{dz}{z}L(z,0)$. Say, the difference between these two values at $\xi = 7$ is within $4$ percent. Therefore, we expect, that the accuracy of the solution given by $A^{(1)}(u) = 1 + \frac{1}{\xi} g(u)$ with $g(u)$ of Eq. (\ref{fitA}) is about $4$ percent for the value of $\xi$ around $\xi \approx 7$. It becomes better with the increase of $\xi$. In general case the expected accuracy of our solution is around $\Big(\frac{3}{2\xi}\Big)^2$. In our approximation the value of $A(0)$ is given by
\begin{eqnarray}
 A(0) &= &A_0(\xi) \approx 1 -\frac{3}{2\xi}  \label{A00}
\end{eqnarray}
As an example we represent the calculated form of the function $A(x)$ for $\xi = 8,9,10$ in Fig. \ref{fig13}.


\section{Asymptotic form of $m(z)$ at $z \gg M^2$}
\label{sectm}

\begin{figure}
\begin{center}
 \epsfig{figure=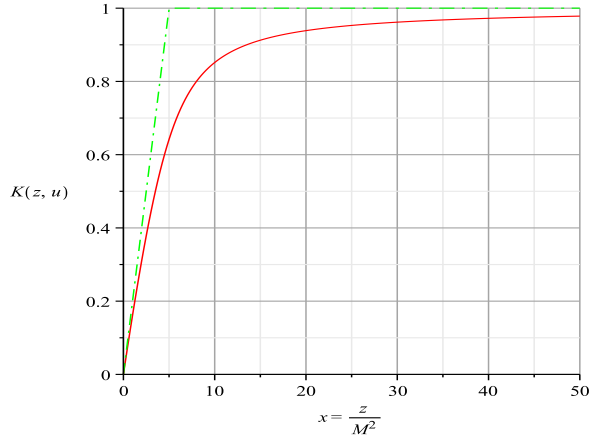,height=60mm,width=80mm,angle=0}
\caption{\label{fig4}  The kernel $K(z,u)$ for $u/M^2 = 5$ as a function of $x=z/M^2$ (solid line) and the approximation via Eq. (\ref{largez}) (dashed - dotted line).    }
\end{center}
\end{figure}

\subsection{Approximation of SD equation by differential equation}

At $u \gg M^2$ we have
\begin{eqnarray}
K(z,u) \approx \frac{{\rm min}(u,z)}{u} = \theta(z-u) + \frac{z}{u} \theta(u-z)\label{largez}
\end{eqnarray}
We combine this with the property $A(z) \approx 1$ at $z\gg M^2$, and approximate the kernel by Eq. (\ref{largez}) at $u > M^2$. As a result we come to the following approximation for the gap equation at $u > M^2$:
\begin{eqnarray}
\frac{\xi}{3} m(u)& \approx&  \int_0^{M^2} {d z} \frac{m(z)A(z)}{A^2(z)(z +  m^2(z))} \frac{z}{u} \label{gapappr} \\&&+ \int_{M^2}^{u} {d z} \frac{m(z)}{(z +  m^2(z))} \frac{z}{u} +  \int_{u}^{\infty} {d z} \frac{m(z)}{(z +  m^2(z))}\nonumber
\end{eqnarray}
This approximation is illustrated by Fig. \ref{fig4}, where the kernel $K(z,u)$ is represented for $u/M^2 = 5$. Eq. (\ref{gapappr}) leads to the following differential equation for $m(z)$:
\begin{equation}
\frac{d^2}{dz^2}\, z \, m(z) + \frac{3}{\xi} \frac{m(z)}{z + m^2(z)} = 0\label{largep}
\end{equation}
with boundary conditions
\begin{eqnarray}
\frac{d}{dz}z\,m(z) &\rightarrow 0 &, \quad z \rightarrow \infty \nonumber\\
z^2 \frac{d}{dz}\,m(z) &\rightarrow -\nu&, \quad z \rightarrow M^2  \label{bound}
\end{eqnarray}
where
\begin{equation}
\nu = \int_0^{M^2}  \frac{z m(z) A(z)d z}{A^2(z)(z +  m^2(z))}
\end{equation}

The solutions of this equation at $z \rightarrow \infty$ behave as
\begin{equation}
m(z) \approx c_+ z^{\frac{-1 + \sqrt{1-\frac{12}{\xi}}}{2}} + c_- z^{\frac{-1 - \sqrt{1-\frac{12}{\xi}}}{2}}\label{solgen}
\end{equation}
with constants $c_{\pm}$. It has been argued in \cite{Miransky:1994vk} that in the theory with $M=0$ the presence of oscillations in the function $m(z)$ is the signature of "falling to the centre" phenomenon. This falling, in turn means that the vacuum should be rearranged and the dynamical chiral symmetry breaking occurs.  For $\xi \ge \xi_{c1} \approx 12$ there is no spontaneous symmetry breaking in the theory while at $\xi < \xi_{c1}$ (that corresponds to $\alpha > \frac{\pi/3}{2 C_2(F)}$) the theory may exist in chirally broken phase. Later this supposition was confirmed by direct lattice simulations. In the latter case the actual dependence of the function $m(z)$ on $z$ may be represented in the form
\begin{equation}
m(z) \approx \, \frac{m(M^2)M}{z^{1/2}} \,\frac{ {\rm sin}\Big({\rm log}\Big(\frac{z^{1/2}}{\mu}\Big)\,\sqrt{\frac{12}{\xi}-1} \Big)}{{\rm sin}\Big({\rm log}\Big(\frac{M}{\mu}\Big)\,\sqrt{\frac{12}{\xi}-1} \Big)},\label{mlz}
\end{equation}
where $\mu$ is the constant factor of the dimension of mass.

In the opposite limit $z \rightarrow 0$ Eq. (\ref{largep}) has the solution
\begin{equation}
m(z) \approx \frac{a}{z} + b
\end{equation}
with constants $a$ and $b$. In the limiting case $M^2=0$ (exchange by massless gauge bosons) the solution should be chosen with $a=0$ (otherwise it is singular at $z =0$). This fixes the value of  $\mu$ in Eq. (\ref{mlz}): $\mu_0 = {\rm exp}(1-2 \, {\rm log}\, 2)\, m(0)$.  Then, parameter $b = m(0)$ becomes equal to the dynamical mass of fermion. In our case of nonzero $M$, we should use the boundary condition of Eq. (\ref{bound}) at $z = M^2$, that gives a certain particular function $\mu_M(\xi)$ of $\xi$ and $M$.

\subsection{Solutions of Eq. (\ref{largep}) with $m \ll M$}

If the finite ultraviolet cutoff is not imposed, the expressions of Eqs. (\ref{solgen}), (\ref{mlz}) satisfy boundary condition Eq. (\ref{bound}) for any values of $\xi$.
However, when the finite ultraviolet cutoff $E_{UV}$ is introduced, the boundary
condition is satisfied only in case $\xi < \xi_{c1}$ and gives
\begin{equation}
{\rm log}\Big(\frac{E_{UV}}{\mu_M(\xi)}\Big)\,\sqrt{\frac{12}{\xi}-1} + {\rm arctg} \sqrt{\frac{12}{\xi}-1} = \pi n \label{cond00}
\end{equation}
 with integer $n$. For the case $M=0$ this leads to the relation between $E_{UV}$ and the dynamical mass $m$ at $\xi \rightarrow \xi_c$ (the so - called Miransky scaling $m(0) \approx 4 E_{UV}\, {\rm exp}\Big(-\frac{\pi n}{\sqrt{\frac{12}{\xi}-1}}\Big)$).  In the case of nonzero $M$ we do not know the particular form of $\mu_M(\xi)$.

Fortunately, Eq. (\ref{gapappr}) is simplified considerably when $m(z) \rightarrow 0$. In this case the expression of Eq. (\ref{mlz}) works in the whole region $M^2 < z$. The value $\nu$ in the boundary condition at $z \rightarrow M^2$ becomes equal to $\nu \approx m(M^2) M^2 \hat{\nu}(\xi)$ with
\begin{equation}
\hat{\nu}(\xi) = \int_0^{M^2}  \frac{A(z)m(z) d z}{M^2m(M^2) A^2(z)} \label{nu0}
\end{equation}
According to the results of the next sections function $B(z) = A(z)m(z)$ varies slowly for $z \in [0,M^2]$. Function $A(z)$ may be estimated using Eqs. (\ref{Afin0}), (\ref{fitA}). This gives $\hat{\nu} \sim 1$.
 Recall that we imply in this section the approximation $A(x) \approx 1$ for $x > 1$. The boundary condition at $z \rightarrow M^2$ gives for such solutions
\begin{equation}
\frac{\mu_M(\xi)}{M} \approx  {\rm exp}\Big(-\frac{\pi n}{\sqrt{\frac{12}{\xi}-1}}\Big) \times {\rm exp}\Big(\frac{{\rm arctg}\,\Big[\frac{1}{2\hat{\nu}(\xi)-1} \sqrt{\frac{12}{\xi}-1}\Big]}{\sqrt{\frac{12}{\xi}-1}}\Big)\label{MSC0}
\end{equation}
with integer $n$. Combining this with Eq. (\ref{cond00}) we get the following algebraic equation for the determination of the values $\xi^{[k]}_{c1}$, $ k = 1,2,...$, at which the solution with small $m(z)$ appears:
\begin{eqnarray}
\pi k &=& \sqrt{\frac{12}{\xi}-1}\, {\rm log} \, \frac{E_{UV}}{M} + \, {\rm arctg}\,  \sqrt{\frac{12}{\xi}-1}\nonumber\\&& - {\rm arctg}\,\Big[\frac{1}{2\hat{\nu}(\xi)-1} \sqrt{\frac{12}{\xi}-1}\Big]\label{MxE}
\end{eqnarray}
Here $k=1$ corresponds to the appearance of the first solution and gives $\xi_{c1}=\xi^{[1]}_{c1} = 12$.
Notice, that this value of critical coupling constant is approximate because Eq. (\ref{gapappr}) is approximate. Actual value of the critical coupling constant may differ somehow from this expression as a consequence of the more complicated form of SD equation with the kernels of Eq. (\ref{kernels}). It is worth mentioning that the similar problem of Eq. (\ref{largep}) with boundary conditions Eq. (\ref{bound}) and with $M\ne 0$ was investigated in \cite{MiranskyInfraredCutoff} for the case $\nu = 0$. When the nonzero cutoff is introduced  at $\xi> \xi_{c1}$ there is no chiral symmetry breaking. When $\xi$  is just below $\xi_{c1}$, then the only solution with dynamical fermion mass appears, and the value $m(M^2)$ is close to zero. When $\alpha$ is increased further (i.e. $\xi$ is decreased further), the dynamical mass becomes large and the additional solutions with smaller masses appear. The number of these additional states is increased with the increase of $\alpha$.
The values of coupling constant $\xi$, at which the new solutions with small $m(z)$ appear are given by the solution of Eq. (\ref{MxE}) that coincides with Eq. (19) of \cite{MiranskyInfraredCutoff} if we substitute $\hat{\nu}(\xi) = 0$. However, only the largest value of dynamical mass corresponds to stable vacuum. This largest value gives rise to the function $m(z)$ that does not have zeros at $z < E_{UV}^2$ and, therefore, corresponds to the main state of the system. The vacua with smaller masses are unstable (tachions appear in the pseudo - scalar channel). We assume that in our case of nonzero $\nu$ the similar pattern takes place.

For the complete SD equation instead of Eq. (\ref{MxE}) the boundary condition at $z = E^2_{UV}$ gives Eq. (\ref{cond00})
with some unknown function $\mu_M(\xi)$. The analysis of the next section will demonstrate, that the actual critical value  $\xi_c$ is smaller than $\xi_{c1}=12$. Notice, that in the analysis of the mentioned solution with small $m(z)$ that just appears at $\xi$ below $\xi_{c1}$ the value $m(M^2)$ in Eq. (\ref{mlz}) remains arbitrary and is not fixed (because in the considered limit Eq. (\ref{largep}) is linear). It will be fixed in the next sections, where we shall consider the numerical solution of the SD equation.

\section{Reduced gap equation}
\label{Sectredgap}

\begin{figure}
\begin{center}
 \epsfig{figure=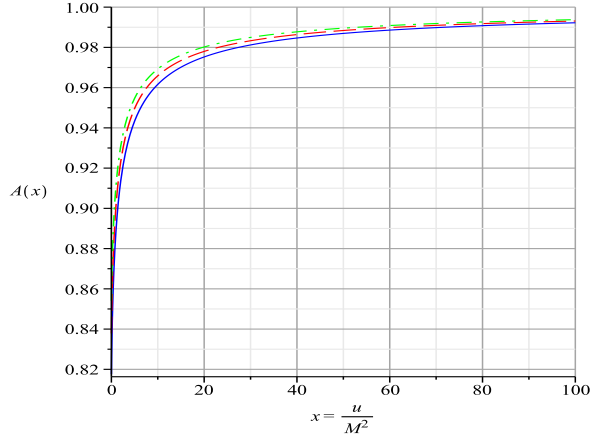,height=60mm,width=80mm,angle=0}
\caption{\label{fig13}  $A(x)$ given by Eq. (\ref{Afin0}), Eq. (\ref{g}) as a function of $x = u/M^2$ for $\xi = 8$ (solid line), for $\xi = 9$ (dashed line), and $\xi = 10$ (dashed - dotted line).     }
\end{center}
\end{figure}

It follows from the above analysis (with the approximate form of the kernel in DS equation Eq. (\ref{gapappr})) that there exists the critical coupling constant at which the solution with small nonzero $m(z)$ appears. The estimated value of this coupling constant obtained in the previous section is $\xi_{c1} = 12$ that corresponds also to the case of pure QED. However, both  Eq. (\ref{gapappr}) and the estimate $\xi_c = 12$ are approximate. In this section the analysis of the SD equation  will allow us to evaluate $\xi_c$ more carefully. Moreover,
the value of $m(M^2)$ in Eq. (\ref{mlz}) was not fixed by the analysis of the previous section (because in the limit $m(z) \rightarrow 0$ Eq. (\ref{largep}) is linear). That's why the actual dependence of the dynamical mass on $\alpha$ was not fixed. This will also be done in the present section.

The first row of Eq. (\ref{gapeq}) may be rewritten as
\begin{eqnarray}
q(y) &=& \frac{3}{\xi} \int_0^{\zeta} {d x} \frac{xq(x)\hat{K}(x,y)}{A^2(x)x +  q^2(x)A^2(0)\hat{m}^2}, \label{gapred00} \\ \hat{K}(x,y)&=& \frac{2}{y+x+1 + \sqrt{(y+x+1)^2-4xy)}}\nonumber
\end{eqnarray}
where we denote $x = \frac{z}{M^2}$, $y = \frac{u}{M^2}$, $\zeta = \frac{E^2_{UV}}{M^2}$, $\hat{m} = \frac{B(0)}{A(0)M}$, and $q(x) = B(z)/B(0)$.  (Notice that $\hat{K}(x,y)$ differs from $K(u,z)$ written for $M=1$ by the factor $1/x$.) Near the critical value of $\xi$ we have $\hat{m} \ll 1$, and the value of dynamical fermion mass is given by $m=\hat{m} M$. In this case we have
\begin{eqnarray}
&&q(y) = \frac{3}{\xi} \int_0^{\zeta} \frac{q(x)dx \hat{K}(x,y)}{A^2(x)} \nonumber\\ &&
-\frac{3}{\xi} \hat{m}^2 \int_0^{\zeta} \frac{q^3(x)A^2(0)dx\hat{K}(x,y)}{A^4(x)(x+\frac{A^2(0)}{A^2(x)}q^2(x)\hat{m}^2)}
 \label{gapred000}
\end{eqnarray}
Function $q(x)$ and the critical value of $\xi$ at which the solution with $m\ne 0$ appears should be calculated via the solution of the homogeneous integral equation
\begin{eqnarray}
&&q(y) = \frac{3}{\xi} \int_0^{\zeta} \frac{q(x)dx \hat{K}(x,y)}{A^2(x)}
 \label{gapredhom}
\end{eqnarray}
This will be done in the next section. Assuming that the solution of the homogeneous equation $q(x)$ and the critical value $\xi_c$ are given we relate the correction $\delta q(x)$ to $q(x)$ to the deviation $\delta \xi$ of $\xi$ from $\xi_c$ and the value of $\hat{m}$:
\begin{eqnarray}
\delta q(y) &\approx & \frac{3}{\xi_c} \int_0^{\zeta} \frac{\delta q(x)dx \hat{K}(x,y)}{A^2(x)}
-\frac{\delta \xi}{\xi_c}q(y) \nonumber\\&& +\frac{3}{\xi_c}\frac{2\delta \xi}{\xi_c^2} \int_0^{\zeta} \frac{q(x)g(x) dx \hat{K}(x,y)}{A^3(x)} \nonumber\\&&-\frac{3}{\xi_c} \hat{m}^2 \int_0^{\zeta} \frac{q^3(x)A^2(0)dx\hat{K}(x,y)}{A^4(x)(x+\frac{A^2(0)}{A^2(x)}q^2(x)\hat{m}^2)}
 \label{gapred1}
\end{eqnarray}
 We multiply both sides of Eq. (\ref{gapred1}) by $q(y)/A^2(y)$ and integrate over $y$:
\begin{eqnarray}
0&=&-\frac{\delta \xi}{\xi_c}\int_0^{\zeta} \frac{dx}{A^2(x)}q^2(x) +\frac{2\delta \xi}{\xi^2_c} \int_0^{\zeta} \frac{q^2(x)g(x) dx }{A^3(x)} \nonumber\\&&- \hat{m}^2 \int_0^{\zeta} \frac{q^4(x)A^2(0)dx}{A^4(x)(x+\frac{A^2(0)}{A^2(x)}q^2(x)\hat{m}^2)}
 \label{gapred2}
\end{eqnarray}
The third integral in the right - hand side is logarithmically divergent at small $x$ for $\hat{m}\rightarrow 0$. Therefore, in the leading order $\sim \hat{m}^2 \, {\rm log}\, \hat{m}^2$ we have
\begin{eqnarray}
-\frac{\delta \alpha}{\alpha_c} \lambda = \frac{\delta \xi}{\xi_c} \lambda &=& -  \frac{m^2}{M^2} \, {\rm log}\, \frac{M^2}{m^2}
,\label{gapeqS}
 \end{eqnarray}
Here we denote
\begin{equation}
\lambda = A^2(0)\Big(\int_0^{\zeta} \frac{dx}{A^2(x)}q^2(x) -\frac{2}{\xi_c } \int_0^{\zeta} \frac{q^2(x)g(x) dx }{A^3(x)}\Big) \label{lambdadef}
\end{equation}
The first integral in the right hand side is logarithmically divergent at $E_{UV} \rightarrow \infty$. Therefore we expect that the value of $\lambda$ grows logarithmically with the increase of $E_{UV}$.

Corrections to Eq. (\ref{gapeqS}) contain the terms $\hat{m}^{2n}{\rm log}^k\hat{m}^2$ with $n>k\ge 0$, and $n=k>1$ that may be neglected compared to $\hat{m}^2 \, {\rm log}\, \hat{m}^2$.
Eq. (\ref{gapeqS}) may be considered as the main result of the present paper. We call it {\it the reduced gap equation}. It is {\it exact} at $m(0) \rightarrow 0$, i.e. it follows from the SD equation without any approximation. All approximations are applied actually during the calculation of parameter $\lambda$.  In principle, $\lambda$ may be considered as the phenomenological parameter.

The dependence of the generated fermion mass on $\alpha$ near its critical value is similar to that of the NJL model:
\begin{eqnarray}
m & \approx & \frac{\sqrt{\lambda_{}\frac{\alpha-\alpha_c}{\alpha_c}}\, M}{\sqrt{\frac{{\rm log}\,{\rm log}\Big({\frac{\alpha_c/\lambda}{\alpha-\alpha_c}}\Big)}{{\rm log}\Big({\frac{\alpha_c/\lambda_{}}{\alpha-\alpha_c}}\Big)} + {\rm log}\,{\rm log}\Big({\frac{\alpha_c/\lambda_{}}{\alpha-\alpha_c}}\Big)+{\rm log}\Big({\frac{\alpha_c/\lambda_{}}{\alpha-\alpha_c}}\Big)}}\label{malpha}
\end{eqnarray}
 Parameter $\lambda$
  does not depend on $C_2(F)$. Therefore, this estimate does not depend on the gauge group.

Eq. (\ref{gapeqS}) may be written in the form that resembles the gap equation of NJL model:
\begin{eqnarray}
{M}^2 &=& \frac{C_2(F) \alpha}{ \pi}\Big({\Lambda}^2 -   {m^2} {\rm log} \Bigl[\frac{{\Lambda}^2}{m^2}\Bigr]\Big),\label{gapeqT}
 \end{eqnarray}
 where
\begin{eqnarray}
{\Lambda}^2(\xi) & = &  \frac{\xi_c}{2} M^2  + \eta (\xi - \xi_c)M^2   \,\label{Lxi}
\end{eqnarray}
 is the new cutoff parameter of the effective NJL model. Here
\begin{equation}
 \eta =  \frac{1}{2} - \frac{\lambda}{\xi_c}
\end{equation}
We used that ${\Lambda}/M$ is of the order of unity and, therefore, ${\rm log} M^2/m^2 \approx {\rm log}{\Lambda}^2/m^2$ for $m\ll M$. The difference of Eq. (\ref{gapeqT}) from that of the NJL model is that the value of the cutoff ${\Lambda}(\xi)$ itself depends on the coupling constant.

\section{Parameters of reduced gap equation}

\subsection{Evaluation of $q(u) = B(u)/B(0)$.}

At the critical value of $\xi$ the SD equation  has the form of Eq. (\ref{gapredhom}).
Function $B(u)$ itself is small. However, the function $q(u) = B(u)/B(0)$ is not small and should be defined as a solution of Eq. (\ref{gapredhom}).  We rewrite it in the form
\begin{eqnarray}
q(y) &=& \frac{3}{\xi} \hat{R}^{} q(y),\label{eqB0} \\
 \hat{R}^{}q(y) &=& \int_0^{\zeta} \frac{d x}{A^2(x)} \frac{2q(x)}{x+y+1+\sqrt{(x+y+1)^2-4xy}}\nonumber
\end{eqnarray}
where $\hat{R}$ is linear operator, $\zeta = \frac{E_{UV}^2}{M^2}$. The solution of this equation may be constructed as a limit of the sequence given by the recursive relation
\begin{equation}
q^{(n+1)}(x) = \hat{R} q^{(n)}(x)/\hat{R}q^{(n)}(0)\label{seq}
\end{equation}
If such a sequence converges to a certain limit $q^{(\infty)}$, this limit gives the solution of Eq. (\ref{eqB0}) that corresponds to the eigenvalue $\xi_c = 3 \, {\rm lim}_{n\rightarrow \infty} \hat{R} q^{(n)}(0)$.
In principle, there exist a lot of different eigenvalues $\xi_c^{[n]}$ that correspond to the solutions $q_{[n]}$ of Eq. (\ref{eqB0}). If the difference of $A(x)$ from unity is neglected, then the solutions that correspond to different eigenvalues will be orthogonal: $\int dx q_{[n]}(x)q_{[m]}(x) = 0$ for $\xi_c^{[n]}\ne \xi_c^{[m]}$. The solutions for $A(x)\ne 1$ may be obtained via continuous deformation of the solutions for $A(x) = 1$.
We are interested in the solution $q(x)$ that corresponds to the largest eigenvalue $\xi_c = \xi_c^{[1]}$.
This is the critical value at which the chiral symmetry breaking occurs.
The corresponding eigenfunction $q(x) = q_{[1]}(x)$ has no zeros.

The asymptotic form of the solution at $x\rightarrow \infty$ is given by Eq. (\ref{mlz}). At the same time for $x \rightarrow 0$ it tends to unity and is analytical in the vicinity of $x=0$.
Therefore, we are looking for the solution in the form that interpolates between the two mentioned limits:
\begin{equation}
q^{}(x) = \frac{{\rm cos}\Big(\frac{1}{2}{\rm log}\Big(1+c_4 x\Big)\,\sqrt{\frac{12}{\xi}-1} \Big)}{(1+ c_1 \, x)^{1/2}}\frac{1+c_2 x + c_5 x^2}{1+c_3 x + c_6 x^2}\label{fitq}
\end{equation}
At a first stage of our calculations we simplify numerical integration using the fit to $q^{(n)}$ of Eq. (\ref{fitq}) at each step. The value of $\xi$ entering $A(x)$ is tuned at each step. We use MAPLE package for the numerical evaluation of integrals. We iterate the sequence until $\xi^{(n)}_c = 3 \, \hat{R} q^{(n)}(0)$ coincides with $\xi^{(n-1)}_c$ within about $0.1$ per cent. Typically, we need $50$ - $100$ iterations for that. If we iterate the sequence further, the value of $\xi^{(n)}_c$ does not tend to any particular limit, but fluctuates instead around a certain value. Those fluctuations are of about $0.2$ per cent of the value. The fluctuations of the calculated values of $\lambda$ are even larger. The reason for this behavior of $\xi_c^{(n)}$ is related to the approximation via Eq. (\ref{fitq}). In order to improve the accuracy of the calculations we need to use the complete space of functions instead of its subset given by the functions of the form of Eq. (\ref{fitq}).

\begin{figure}
\begin{center}
 \epsfig{figure=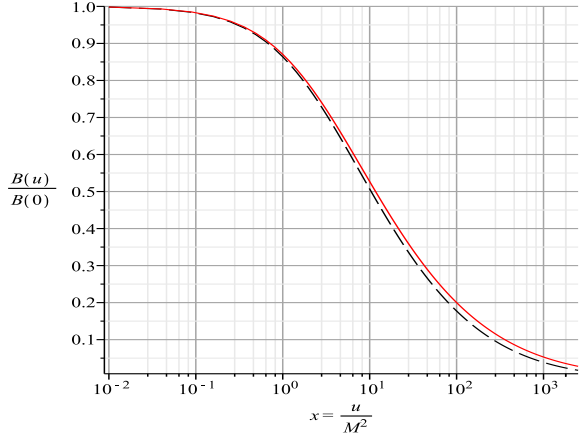,height=60mm,width=80mm,angle=0}
\caption{\label{fig17} Solution $q(x)$ of  Eq. (\ref{eqB0})  for $E_{UV} = 100 M$ (solid line) and $E_{UV} = 50 M$ (dashed line).    }
\end{center}
\end{figure}

Our next step is to use the more general form of $q^{(n)}(x)$ to refine the value of $\xi_c$ and the shape of $q(x)$.
Instead of Eq. (\ref{fitq}) we apply spline approximation at each step of the iteration procedure. Namely, the integral over $x$ in Eq. (\ref{eqB0}) is evaluated at $91$ points $y$ within the interval $x \in [0, \zeta]$ (located with density $\sim 1/x^p$, where $p$ is tuned accordingly). The resulting function is interpolated by degree $3$ splines. The result is substituted again into Eq. (\ref{seq}). This procedure is repeated iteratively. It improves the estimates of the values of $\xi_c$, $\lambda$ and the shape of $q(x)$. The corrections are negligible for $E_{UV}\le 50 M$ and increase with the increase of $E_{UV}$. The corrections to the values of $\xi_c$ remain within about $1$ per cent. The corrections to the values of $\lambda$ are larger. The especially valuable correction (of about $10$ per cent) is received by the value of $\lambda$ at  $E_{UV} = 100 M$.  We found no additional corrections to our results using the larger number of points ($161$ instead of $91$). Our final  results are represented in Table \ref{tab.01}.
As an example, we represent the shapes of $q(x)$ for $E_{UV} = 50 M, 100 M$ in Fig. \ref{fig17}.

\begin{table}
\begin{center}
\begin{tabular}{|c|l|l|l|l|l|l|}
  \hline
  $\frac{E_{UV}}{M}$ & 5 & 10 & 20 & 30 & 40 & 50  \\ \hline
  $\xi_c$ & $6.83$ & $7.89$ & $8.69$ & $9.06$ & $9.29$ & $9.44$   \\
  \hline
  $\lambda$ & $3.6$ & $5.9$ & $9.0$ & $11.1$ & $12.9$ & $14.4$  \\
  \hline\hline
  $\frac{E_{UV}}{M}$ &  60 & 70 & 80 & 90 & 100 &\\ \hline
  $\xi_c$ &  $9.57$ &$9.66$ & $9.74$ & $9.81$ & $9.88$ & \\
  \hline
  $\lambda$ & $15.7$ & $16.8$ & $17.8$ & $18.7$ & $19.6$&   \\
  \hline
\end{tabular}
\caption{\label{tab.01} The critical coupling constant $\xi_c$ and parameter $\lambda$ as the functions of the cutoff. Error bars correspond to the last digits of presented values. For example $6.83$ means $6.83\pm 0.01$.  The given error bars ignore that  Eq. (\ref{Afin0}) and Eq. (\ref{fitA}) are approximate. The accuracy in the determination of $A(x)$ via Eqs. (\ref{Afin0}), (\ref{fitA}) induces the error bars of about $4$ per cent for all calculated quantities.}
\end{center}
\end{table}

\subsection{Evaluation of $\lambda$ and $\eta$.}

Parameter $\lambda$ entering Eq. (\ref{gapeqS}) is given by Eq. (\ref{lambdadef})
 The results of our calculations are presented in Table \ref{tab.01}.
One can see, that the value of $\lambda$ grows with the increase of $E_{UV}$ as expected.
This indicates that in the limit $E_{UV} \rightarrow \infty$ we already cannot have the second order phase transition because infinitely small variation of the coupling constant causes the finite change of dynamical fermion mass. Therefore, this limit in the given system, presumably, corresponds to the first order phase transition with the discontinuous change of observables.

The value of $\eta = \frac{1}{2} - \frac{\lambda}{\xi_c}$ entering Eq. (\ref{Lxi})  varies between $\eta \approx -0.03$ for $E_{UV} = 5 M$ and $\eta = -1.48$ for $E_{UV} = 100 M$.

\subsection{Evaluation of critical coupling constant}

\begin{figure}
\begin{center}
 \epsfig{figure=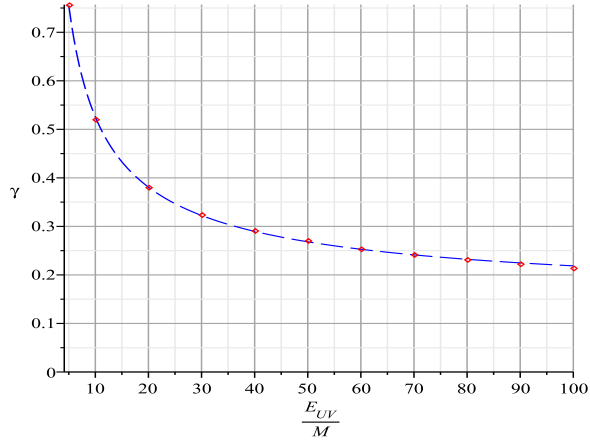,height=60mm,width=80mm,angle=0}
\caption{\label{fig22} The value of $\gamma$ as a function of the cutoff (points) and our fit $\gamma \approx 0.134141+1.813487\, \Big(\frac{M}{E_{UV}}\Big)^{0.665774}$ (dashed line).}
\end{center}
\end{figure}

We represent the value of critical coupling constant in the form
\begin{equation}
\alpha_c \approx (1+\gamma)\frac{1}{2C_2(F)}\frac{\pi}{3},\label{crit2}
\end{equation}
where parameter $\gamma$ depends on the ratio $E_{UV}/M$ and is related to the critical value $\xi$ as
\begin{equation}
\gamma = \frac{12}{\xi_c} - 1
\end{equation}
At $\alpha > \alpha_c$ the gap equation has the nontrivial solution $m\ne 0$.
We represent the value of $\gamma$ as a function of $E_{UV}$ in Fig. \ref{fig22}.
One can see, that the calculated values $\alpha_c$ remain larger than  $\alpha_{c1} = \frac{\pi/3}{2 C_2(F)}$. At the same time $\gamma$ is decreased slowly with the increase of $E_{UV}$. In principle, we do not exclude at the present moment that ${\rm lim}_{E_{UV}\rightarrow \infty} \gamma = 0$. However, the detailed analysis of the system at large values of cutoff is needed to check this possibility.

\subsection{Error bars}

\label{bars}

There are three sources of error bars in the quantities calculated in the present section:
\begin{enumerate}

\item{}
The error bars in the procedure discussed in the present section originate from the use of numerical methods for the solution of integral equation. We achieve in our numerical method described above the accuracy less than $0.2$ per cent for the values of $\xi_c$. The error bars for the values of $\lambda$ are larger (see Table \ref{tab.01}). In principle, this accuracy may be improved extending the number of iterations and using more refined discretization. However, this is not reasonable because of the error bars originated from the determination of function $A(x)$.

\item{}

There is the error bar that originates from the approximation of $g(x)$ of Eq. (\ref{Afin0}) via Eq. (\ref{fitA}). However, as demonstrated in Fig. \ref{fig3} the actual difference between Eq. (\ref{Afin0}) and Eq. (\ref{fitA}) is, at least, smaller than the resolution of the plot. Therefore we neglect the corresponding error bar.

\item{}

The largest error bar originates from the uncertainty in the determination of $A(x)$ using Eq. (\ref{Afin0}). The accuracy of this approximation is better than  $\Big(\frac{3}{2\xi}\Big)^2$ that is about $4$ per cent for $\xi > 7$.

\end{enumerate}

Overall, the expected accuracy of the numerical estimates given here is within $4$ per cent.

\section{Discussion and conclusions}

Eq. (\ref{gapeqT}) appears as the one - loop gap equation in the NJL model with the four - fermion coupling constant $\kappa = \frac{4 \pi \alpha}{M^2}\times \frac{2C_2(F)}{N}$ and the cutoff ${\Lambda}^2 \approx \frac{\pi}{\alpha_c  C_2(F)} M^2 + \frac{2\pi\eta}{C_2(F)} (\frac{1}{\alpha} - \frac{1}{\alpha_c})M^2  + ...$ The value of critical coupling constant $\alpha_c$ and the value of $\eta$ were calculated in the previous section. However, they may also, in principle, be considered as the phenomenological parameters. The corresponding effective action is given by
\begin{equation}
S_{\rm eff} =  \int d^4 x  \, \Bigl[i\bar{\psi} \gamma \partial \psi + \kappa
 (\bar{\psi}_L \psi_R)(\bar{\psi}_R \psi_L) \Bigr]\label{NJL02}
\end{equation}
The value of the cutoff to be used in this effective NJL model also depends on $\alpha$. When $\alpha$ approaches its critical value, the value of $\kappa$ approaches its critical value that follows from the one - loop gap equation of the NJL model. It is worth mentioning that our results may easily be generalized to the wide class of gauge groups and fermion representations. In practise, such a generalization results in the substitution $C_2(F)\rightarrow C_2(R)$, where $C_2(R)$ is the eigenvalue of quadratic Casimir operator of the gauge group $G$ for the representation $R$, in which the fermions $\psi$ are arranged.
Notice, that near to the critical value of $\alpha$ the gap equation of the NJL model given by Eq. (\ref{NJL02}) is {\it exactly} equivalent to Eq. (\ref{gapeqS}) that follows from the SD equations.

Although the dynamical fermion mass can be calculated using the effective NJL model of Eq. (\ref{NJL02}), the whole fermion propagator differs from that of the NJL model. Namely, the propagator in Euclidean space - time   is given by
\begin{equation}
D(p) = \frac{1}{A(p^2)\Big(p\gamma - i m(p^2) \Big)},
\end{equation}
where function $A$ is given by Eq. (\ref{Afin0}). It tends to unity at $p^2/M^2\rightarrow \infty$, and differs essentially from unity at $p^2/M^2 \ll 1$. (NJL model predicts $A(p^2)=1$). Function $m(p^2)$ is the running fermion mass. Near criticality its value at $p^2=0$ gives the value of the pole mass. Function $q(p^2) = B(p^2)/B(0) = \frac{m(p^2)A(p^2)}{m(0)A(0)}$ above $p^2 = M^2$  is decreased as $q(p^2) \sim \frac{M}{p}$. For $0<p^2<M^2$ it is decreased by about $15$ per cent. Therefore, the NJL approximation of Eq. (\ref{NJL02}) allows to evaluate the fermion propagator only for momenta $p^2 \ll M^2$ (then $A(0)$ may be considered as the wave function renormalization constant). The forms of functions $A(x)$ and $q(p^2)$ are represented in Fig. \ref{fig13} and Fig. \ref{fig17}.

Our estimate for the critical value $\alpha_c$ depends on the value of cutoff $E_{UV}$. It is given by $\alpha_c = (1+\gamma)\frac{1}{2C_2(F)} \frac{\pi}{3}$ with $\gamma$ that decreases slowly with the increase of $E_{UV}$. For $E_{UV} = 10 M$ it is about $\gamma \approx 0.5$ while for $E_{UV}=100 M$ it is about $0.2$. For the considered values of $E_{UV}$  the value of $\alpha_c$ is larger than that of the case $M=0$ given by $\alpha^{M=0}_c = \alpha_{c1} = \frac{1}{2C_2(F)} \frac{\pi}{3}$. Our analysis shows that $\alpha_c$ becomes closer to $\alpha_{c1}$ when the value of $E_{UV}$ is increased. We do not exclude, that this value tends to $\alpha_{c1}$ at $E_{UV} \rightarrow \infty$. However, the check of this possibility requires an additional investigation. In principle, we may, to a certain extent, take into account the contributions that were disregarded in the ladder approximation via a change of the function $q(p^2)$ at $p^2 > M^2$.
In this extended approach $\alpha_c$ is already not resulted solely from the ladder SD equation, and becomes the phenomenological parameter. However, we expect that it remains of the order of its values calculated here.
Anyway, the ladder Dyson - Schwinger equation itself is a rough approximation to the whole theory. We expect that our analysis gives the correct qualitative pattern of the chiral symmetry breaking in the given system, and more or less reasonable approximation for the dependence of the dynamical fermion mass on $\alpha - \alpha_c$. The latter dependence is given by the NJL approximation with the parameters discussed above.

The calculated values of critical coupling constant and the constant $\lambda$ of Eq. (\ref{gapeqS}) depend on ultraviolet cutoff $E_{UV}$. Actually, the appearance of such a cutoff is very natural. It may be understood if we take into account that in the original renormalizable gauge theory the value of running coupling constant $\alpha(p^2)$ depends on momentum. This dependence becomes relevant for the momenta above a certain scale that may be identified with $E_{UV}$. For the theory given by Eq. (\ref{S0}) this scale is of the order of the scalar boson mass. Above this scale vector bosons may be considered as massless. When the gauge theory is asymptotic free, this will result in the falling of the function $m(p^2)$ as $1/p^2$  at $p > E_{UV}$ that is faster than $1/p$ obtained here \cite{MiranskyQCD,Miransky:1994vk}. This results effectively in the appearance of the ultraviolet cutoff in our expressions. We assume, that $E_{UV} \gg M$.

At not very large values of the ratio $E_{UV}/M$ the Dyson - Schwinger equation predicts the second order phase transition of the NJL type between the symmetric and the broken phases. Parameter $\lambda$ of Eq. (\ref{gapeqS}) grows when this ratio is increased. That means that the strength of the phase transition is increased. Formally, at the infinite value of $E_{UV}/M$ we already do not have the second order phase transition because infinitely small variation of the coupling constant near its critical value causes the finite change of dynamical mass. This indicates that in the complete theory the second order phase transition may become the first order phase transition at finite but sufficiently large ratio $E_{UV}/M$.

The model considered in the present paper is to a certain extent similar to that of \cite{MiranskyInfraredCutoff}, where the system with exchange by massless gauge bosons was considered in the presence of infrared cutoff. This infrared cutoff plays the role similar to the role of vector boson mass in our model. Namely, when coupling constant $\alpha$ is just above the critical value $\alpha_c$, the solution of Dyson - Schwinger equation with nonzero but small dynamical mass of the fermion appears. When $\alpha$ is increased further, the mass of the first solution is increased, and just above a certain value $\alpha_c^{(2)}$ the second solution appears, etc. Only the solution with the largest value of mass is stable. The other solutions correspond to unstable vacua. However, those solutions are in one - to one correspondence with the resonances in the pseudoscalar channel \cite{Miransky:1981rt}. Therefore, we expect the appearance of such resonances in our case as well. In the case of massless gauge bosons these resonances are light (their masses are of the order of the fermion mass), and, therefore, they may become light in our case as well. However, examination of this hypothesis requires analytical investigation of Bethe - Salpeter equations and/or lattice numerical simulations, and is out of the scope of the present paper.

The weak coupling expansion does not work near to the critical value of $\alpha$ because this critical value multiplied by $C_2(F)$ is not small. However, our observation that this product remains of the order of unity matches the t'Hooft condition necessary for the application of large $N$ expansion in gauge theory \cite{largeN}
\begin{equation}
\frac{g^2}{4\pi} N \sim 1 \label{thoft}
\end{equation}
Direct application of the large $N$ expansion in gauge theory is a rather complicated task. However, in QCD typically the large $N$ expansion is applied to the effective models \cite{largeNphenom}. We suppose, that the same may be done in our case as well. We dealt with two approximations: the one defined by ladder Schwinger - Dyson equation, and the NJL approximation. The relation between the two has been discussed above. In the NJL model the $1/N$ expansion is a reasonable approximation, and, actually, is the only one that may be successfully applied. The problem with the next to leading order contributions mentioned in the Introduction, most likely, is not relevant here. Because the phase transition is likely of the second order (that is confirmed by the direct lattice investigation for the case of $SU(2)$ and $U(1)$ groups \cite{SU2lattice,U1lattice}), the dynamical fermion mass near to the phase transition is much smaller than the mass of vector boson. This may occur in NJL model {\it only} if it is defined with the additional counter - terms that cancel quadratic divergencies in the next - to leading order expressions (or, equivalently, is defined in zeta/dimensional regularization with suitably redefined four - fermion coupling, - see again the discussion in Introduction). Therefore, we suppose, that the actual NJL approximation to the considered theory is of this type, and the $1/N$ expansion works in it as a result. This is in accordance with the observation that in the original renormalizable theory given by the action of Eq. (\ref{S0}) the  t'Hooft condition Eq. (\ref{thoft}) is satisfied near to the criticality.

The important feature of the NJL model that was not considered here is the Nambu sum rule \cite{VZ2012,VZ2013,Z2013JHEP,VZ2013_2} that relates the values of masses of scalar excitations with the value of dynamical fermion mass. In our case this sum rule gives for the Higgs boson $M_H = 2 m$. The confirmation of this expression would become the additional justification of the hypothesis that the NJL model of Eq. (\ref{NJL02}) approximates the model of Eq. (\ref{S0}). Partially, this work has been done already in the series of papers on Bethe - Salpeter (BS) equations and on the light front dynamics of bound states in gauge theory. Actually, the works on light front dynamics indeed confirm the relation $M_H \approx 2 m$ \cite{lightcone} in the system with the exchange by massive vector bosons near to the criticality. We are aware of the investigations in the approach with BS equations only for the case $M=0$ (see, for example, \cite{Dorkin:2007wb,Miransky:1994vk} and references therein), where this relation is confirmed as well.

Finally, we would like to notice the possible application of the obtained results to the construction of various models of composite Higgs bosons. At the present moment such models are developed actively and in many cases they are based on  various versions of NJL model. However, the problems with the NJL models defined in conventional cutoff regularization that were mentioned in the Introduction make many of such constructions questionalble. In fact, most of the physicists working on this subject have in mind that there is a renormalizable theory behind the NJL model, that improves it considerably. The present work demonstrates that indeed the massive gauge theory with the working scale $E_{\rm high}$ and the gauge boson mass $M$ may stand behind the effective low energy NJL model that gives rise to dynamical fermion mass $m \ll E_{\rm high}, M$. The dangerous quadratic divergences of such effective NJL model are, presumably, cancelled by the high energy renormalizable gauge theory, which allows to consider seriously the NJL approximation with the four - fermion interactions. Actually, the justification of the use of NJL approximation was the starting point of the present investigation.

The author is greatful to V.A.Miransky for numerous discussions and for the explanation of various aspects of Schwinger - Dyson equations in gauge models.
The work  is  supported by the Natural Sciences and Engineering Research Council of
Canada.

\end{document}